\documentclass[final,3p,times,twocolumn]{elsarticle}

\usepackage{graphicx}
\usepackage{epstopdf}

\usepackage{amssymb}

\biboptions{square,sort&compress}

\journal{Optics Communications}

\begin{document}
\begin{frontmatter}



\title{Super/Subradiant Second Harmonic Generation}


\author{Gennady A. Koganov\corref{cor1}}
\ead{quant@bgu.ac.il}

\author{Reuben Shuker}
\ead{shuker@bgu.ac.il}

\cortext[cor1]{Corresponding author: Gennady Koganov}

\address{Physics Department, Ben-Gurion University of the Negev, \\ P.O.B. 653, Beer Sheva, 84105, Israel}

\begin{abstract}
A scheme for active second harmonics generation is suggested. The system comprises $N$ three-level atoms in ladder configuration, situated into resonant cavity. It is found that the system can lase in either superradiant or subradiant regime, depending on the number of atoms $N$. When $N$ passes some critical value the transition from the super to subradiance occurs in a phase-transition-like manner. Stability study of the steady state supports this conclusion.
\end{abstract}


\end{frontmatter}

\section{Introduction}

The first model of statioanry superradiance, a superradiant laser was described by Haake et. al. \cite{Haake93}, who considered a model of three-level atoms placed inside a resonant cavity, and pumped with a classical external electromagnetic field. In addition, another "passive" cavity mode was used to coherently couple  one of the non-lasing atomic transition. This "passive" cavity mode, being adiabatically eliminated, results in nonlinear collective decay. The steady state laser intensity calculated for this model scales as $N^2$ typical for superradiance. The superradiance obtained in Ref. \cite{Haake93} essentially differs from the Dicke superradiance\cite{Dicke} by: (i) it is stationary rather than transient, (ii) the linewidth of the superradiant laser scales as $1/N^2$ which is extremely small compared to the spectral width of Dicke superradiant pulse, (iii) intensity fluctuations of the superradiant laser are essentially squeezed, while those of superfluorescence pulse are close to fluctuations of a coherent state. Since then  many theoretical works have been devoted to study this model as well as some other models \cite{Haake96,Haake96-2,Golubev,Chen,Meiser-2010,Vogl2011,Gerace-2011,Bohnet2012}.

The key mechanism responsible for stationary superradiance in such lasers is collective nonlinear spontaneous decay of one of the atomic states. Such a cooperative decay is provided by incorporating an additional ''passive'' resonator to couple the non-lasing atomic transition  \cite{Haake93,Haake96}, or to couple two non-lasing transitions in a four-level scheme \cite{Chen}. Additionally, in order to reach the effect of super-radiance, when the laser intensity scales as $N^2$, the pumping field strength was taken proportional to $N$. However, as we have shown in \cite{OurOptLett-2011}, the two aforementioned requirements are not necessary to obtain superradiant lasing. We proposed a model of superradiant laser based on $N$ three-level systems in ladder configuration, driven by two pumping lasers. All spontaneous decay processes are linear, i.e. no correlation in spontaneous emission is introduced. We called this new type of superradiance Field Driven Superradiance since it stems from simultaneous coherent interaction of the atomic system with two driving laser fields. It was found that in the steady state, the generated laser field phase is locked to the relative phase of the  two pumping lasers, and at a strong enough pump, the number of photons inside the resonator scales as $N^2$. It was also found that this system exhibits subradiant behavior manifested in a departure from $N^2$ scaling, in appropriate conditions.\\

In this paper we suggest a scheme of active super/subradiant second harmonics generation in semiconductor quantum wells (SQW) \cite{PhysRevLett.84.1019,Gmachl}. Within the scope of this article we consider $N^2$ dependence of the laser intensity as an indication of stationary superradiance, whereas saturation with $N$ as indication of subradiance. Utilizing semiclassical treatment, we solve optical Maxwell–Bloch equations both numerically and, in some particular cases, analytically. It is shown that the second harmonics generated by the system can exhibit both superradiance and subradiance, depending on the number of atoms, which is strong manifestation of collective behavior. The phase of the generated field is locked to that of the pumping laser. We have found that in the vicinity of some critical value of the pump Rabi frequency the system behaves in a phase-transition-like manner, changing from super to subradiant regime. \\

\section{Problem formulation}

\begin{figure}[h]
\begin{center}
\includegraphics[totalheight=0.25\textheight,clip=true]{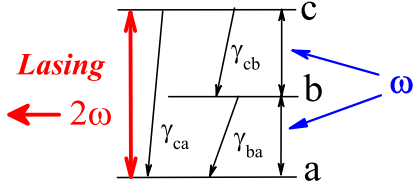}
\caption{(Color online) Schematic diagram of three-level system for second harmonics generation.}\label{scheme}
\end{center}
\end{figure}

Consider a three-level system shown in Fig. \ref{scheme}. The two transitions $\left|a\right\rangle\rightarrow\left|b\right\rangle$ and $\left|b\right\rangle\rightarrow\left|c\right\rangle$ have closed enough frequencies, i.e. $\omega_{ab}\approx \omega_{bc}$, to the extent that both transitions can be driven by a single pumping laser. The transition $\left|a\right\rangle\rightarrow\left|c\right\rangle$ is in resonance with the laser resonator, therefore the frequency of the generated laser field will be twice the frequency of the driving laser. Relaxation constants $\gamma_{ij}$ describe the linear spontaneous decay of the corresponding atomic states, where the events of spontaneous emission of different atoms are independent of each other. The interaction Hamiltonian for a single atom is given by

\begin{equation}\label{Hamiltonian}
\vbox{$H_{int}=-\hbar[\Omega e^{-i\varphi}(\sigma_{ab}e^{-i\Delta t}+\sigma_{bc}e^{i\Delta t})+g\hat{b} e^{i\Delta_{ac}t}\sigma_{ac}]$ + H.c.}
\end{equation}

\noindent where $\sigma_{ij}=\left|j\rangle\langle i\right|$ are the atomic raising/lowering operators,  $\Omega =\left|\vec{d}\cdot\vec{E}\right|/\hbar$ is Rabi frequency of the pumping laser, $\varphi$ is the phase of the pumping field, $\vec{d}$ is the relevant dipole matrix element, 
$\vec{E}$ is the pumping field amplitude,  $\Delta$ ($-\Delta$) is the detuning of the pumping field from the atomic transition $\left|a\right\rangle\rightarrow\left|b\right\rangle$ ($\left|b\right\rangle\rightarrow\left|c\right\rangle$), $\Delta_{ac}$ is the detuning of the transition $\left|a\right\rangle\rightarrow\left|c\right\rangle$ 
from the resonator mode frequency, g is the coupling constant and $\hat{b}$ is the photon annihilation operator. \\

Semiclassical equations of motion derived from the master equation with Hamiltonian (\ref{Hamiltonian}) read:

\begin{equation}\label{rcc}
\dot{\rho}_{cc}=-\left(\gamma_{cb}+\gamma_{ca}\right)\rho_{cc}+
i\Omega(\rho_{cb}e^{i\varphi}-\rho_{bc}e^{-i\varphi})
+ig^{*} b\rho_{ca}-ig b^{\dagger}\rho_{ac}
\end{equation}
\begin{equation}\label{rbb}
\dot{\rho}_{bb}=\gamma_{cb}\rho_{cc}-\gamma_{ba}\rho_{bb}+i\Omega(\rho_{bc}e^{-i\varphi}- \rho_{cb}e^{i\varphi}-\rho_{ab}e^{-i\varphi}+\rho_{ba}e^{i\varphi})
\end{equation}
\begin{equation}\label{rbc}
\dot{\rho}_{bc}=-(\frac{\gamma_{ca}+\gamma_{cb}+\gamma_{ba}}{2}-i\Delta)\rho_{bc}-i\Omega[(\rho_{cc}-\rho_{bb})e^{i\varphi}+\rho_{ac}e^{-i\varphi}]+ig^*b^{\dagger}\rho_{ba}
\end{equation}
\begin{equation}\label{rac}
\dot{\rho}_{ac}=-\frac{\gamma_{ca}+\gamma_{cb}}{2}\rho_{ac}-ig^{*}b^{\dagger}(\rho_{cc}-\rho_{aa})+i\Omega e^{i\varphi}(\rho_{ab}-\rho_{bc})
\end{equation}
\begin{equation}\label{rab}
\dot{\rho}_{ab} =-(\frac{\gamma_{ba}}{2}+i\Delta)\rho_{ab}-i\Omega[(\rho_{bb}-\rho_{aa})e^{i\varphi}-\rho_{ac}e^{-i\varphi}]- ig^{*}b^{\dagger}\rho_{cb}
\end{equation}
\begin{equation}\label{b}
\dot{b}=-\kappa b+i\sum_{j}^{N}g^{*}\rho^{j}_{ca}
\end{equation}

\noindent where $N$ is the number of atoms. \\
This is a standard set of optical Maxwell-Bloch equations where irreversible processes of spontaneous emission and resonator losses are governed by linear terms with atomic relaxation constants $\gamma_{ij}$ and the field decay rate $\kappa$ in the cavity. In semiclassical approximation the field operators can be replaced with c-numbers, so we put $b=\sqrt{n}e^{i\Psi}$, where n is the number of photons in the lasing mode and $\Psi$ is the laser field phase. No assumption is made regarding the initial atomic cooperativity such as nonlinear collective relaxation imposed by the presence of a ''passive'' resonator, as in \cite{Haake93,Haake96,Chen}.  \\

\section{Results and discussion}

We have solved Eqs. (\ref{rcc})-(\ref{b}) both numerically and, in particular case of resonance analytically using proper approximations. At resonance $\Delta_{ac}=\Delta=0$ steady state solution of Eqs. (\ref{rcc})-(\ref{b}) can be obtained analytically. For the sake of simplicity we put three atomic relaxation rates equal, i.e.  $\gamma_{ij}=\gamma$. Among nine formal solutions for the photon number $n$ the two physically acceptable solutions ($n>0$) can be approximated, using $g\ll \Omega$, as follows: 

\begin{equation}
n_1\approx \frac{4 g^2 \gamma^4 \Omega^4}{\kappa^2 (\gamma^4 + 
   12 \gamma^2 \Omega^2 + 12 \Omega^4)^2}N^2,
   \label{n1}
\end{equation}

\begin{equation}
n_2\approx \frac{4 \Omega^2 - 3\gamma^2}{4 g^2}.
   \label{n2}
\end{equation}

\begin{figure}[h]
\begin{center}
\includegraphics[totalheight=0.25\textheight,clip=true]{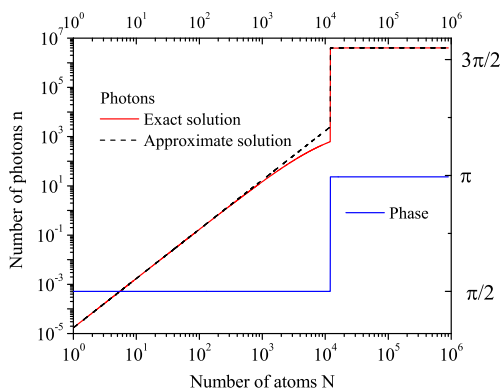}
\caption{(Color online) Number of photons and the field phase as a function of the number of atoms.}\label{n(N)}
\end{center}
\end{figure}

Expression (\ref{n1}) is valid at $N\ll N_c$, while expression (\ref{n2}) is valid for $N>N_c$, where

\begin{equation}
N_c\approx \frac{3 \kappa \Omega^2}{\gamma g^2}.
   \label{Nc}
\end{equation}

\noindent is a critical number of atoms that separates the superradiant regime from the subradiant one, as will be seen in the following. The field phase $\Psi\approx 2\varphi+\pi/2$ for $N<N_c$ and $\Psi \approx 2\varphi+ArcTan[\gamma/\Omega]$ for $N>N_c$. In Fig. \ref{n(N)} the number of photons and the field phase are plotted as a function of the number of atoms. One can see that approximate analytical Eqs. (\ref{n1}) and (\ref{n2}) describe quite well the system behavior, except the vicinity of the critical point $N\sim N_c$, where it differs from the numerical solution.

\begin{figure}[h]
\begin{center}
\includegraphics[totalheight=0.25\textheight,clip=true]{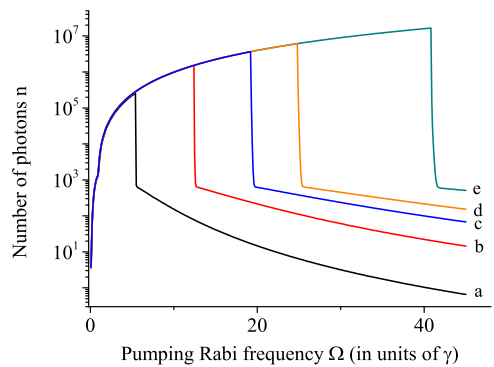}
\caption{(Color online) Number of photons as a function of pumping Rabi frequency at various values of the number of atoms N: a)1000, b) 5000, c) 12000, d) 20000, e) 50000.}\label{n(Om)}
\end{center}
\end{figure}

\begin{figure}[h]
\begin{center}
\includegraphics[totalheight=0.25\textheight,clip=true]{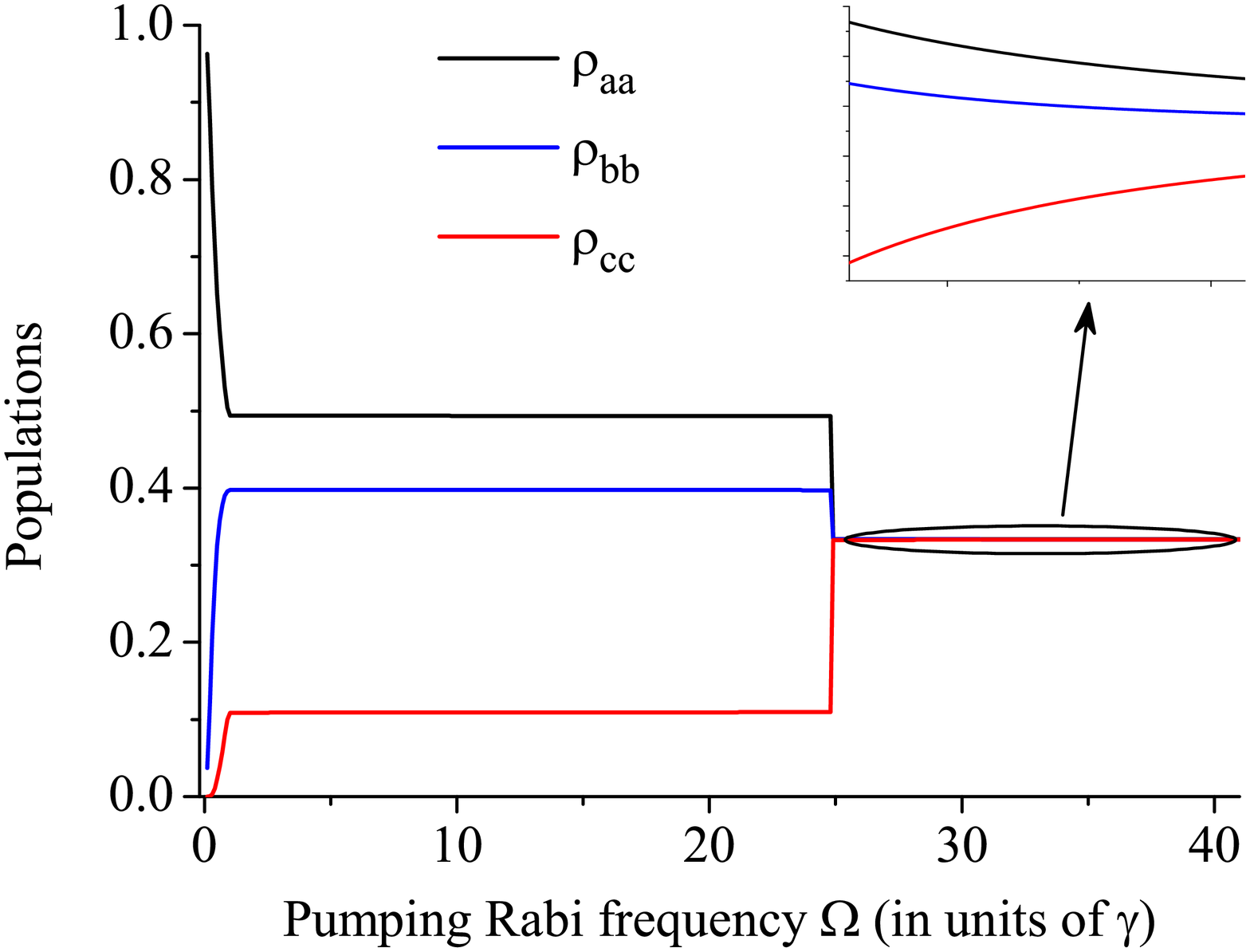}
\caption{(Color online) Populations as a function of pumping Rabi frequency at $N=20000$. The inset shows convergence of atomic populations to 1/3.}\label{Rho(Om)}
\end{center}
\end{figure}

\begin{figure}[h]
\begin{center}
\includegraphics[totalheight=0.25\textheight,clip=true]{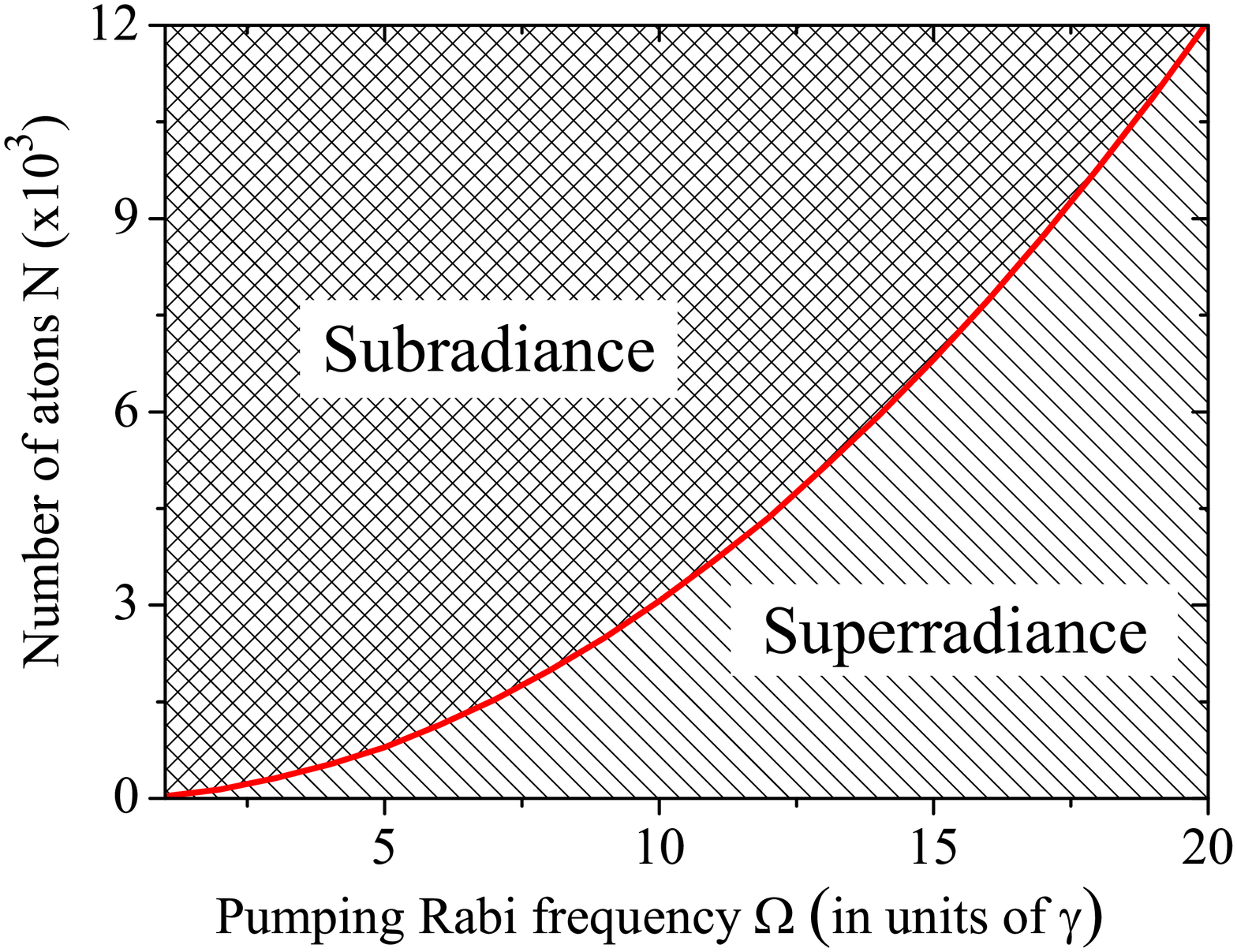}
\caption{(Color online) Phase diagram $n$ vs $\Omega$.}\label{PhaseDiag-Super}
\end{center}
\end{figure}

\begin{figure}[h]
\begin{center}
\includegraphics[scale=0.25,clip=true]{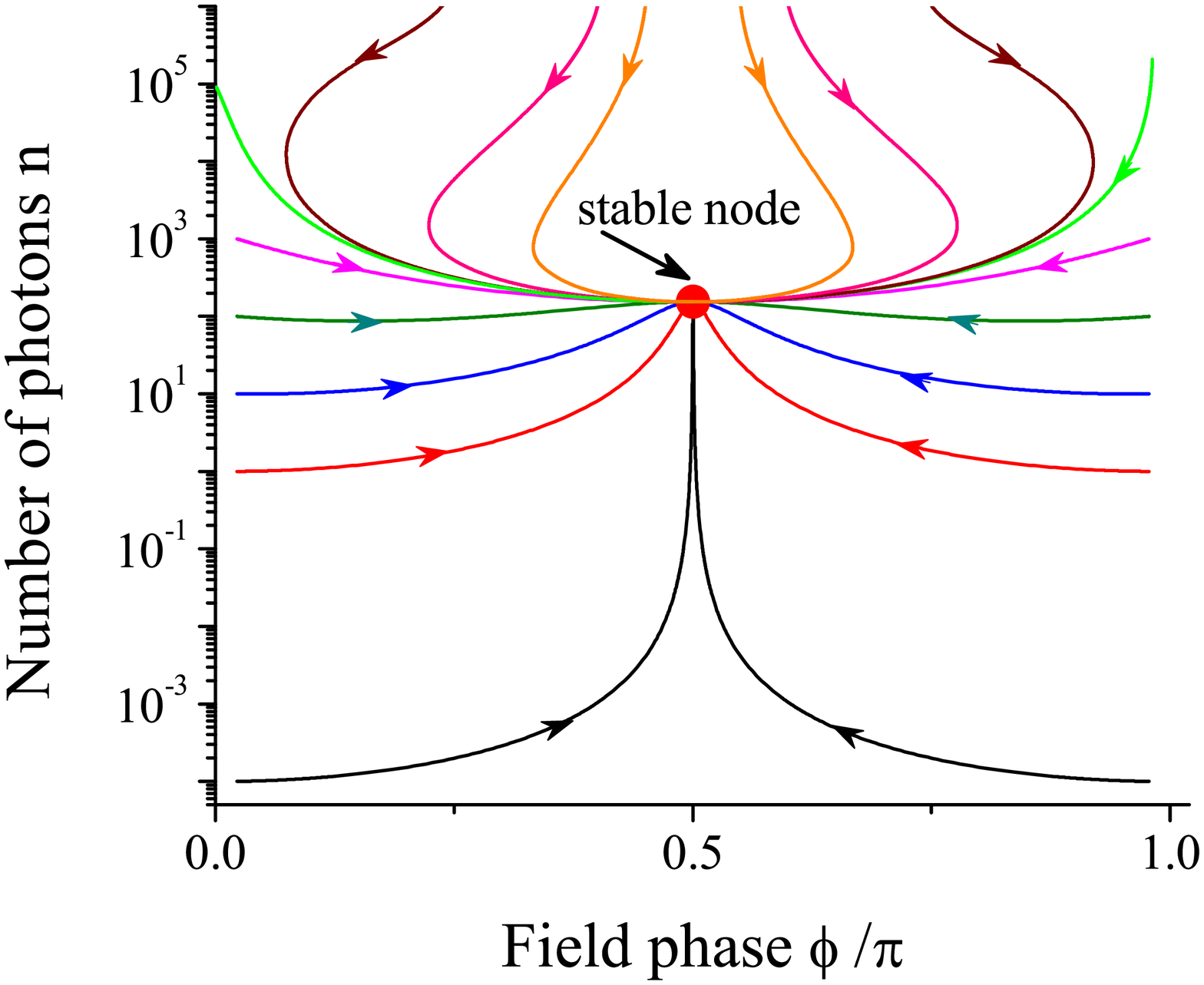}
\caption{(Color online) Superradiant regime: phase portrait of $n(t)$ vs $\phi(t)$.}\label{Stab-Super}
\end{center}
\end{figure}

\begin{figure}[h]
\begin{center}
\includegraphics[totalheight=0.25\textheight,clip=true]{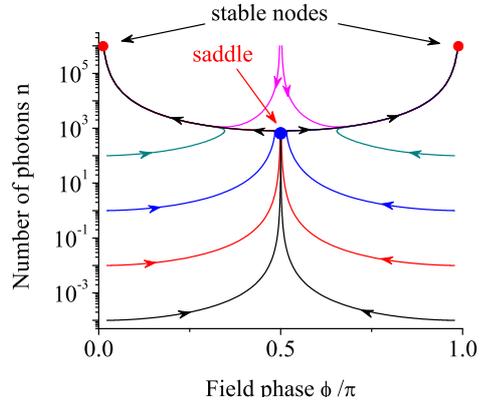}
\caption{(Color online) Subradiant regime: phase portrait of $n(t)$ vs $\phi(t)$.}\label{Stab-Sub}
\end{center}
\end{figure}

When $N$ approaches its critical value $N_c$ defined by Eq. (\ref{Nc}), both the number of photons and the field phase abruptly change their values, which indicates to the presence of a phase transition. At $N\ll N_c$ the photon number scales as $N^2$ in agreement with Eq. (\ref{n1}) - a signature of superradiance. At $N=N_c$ the photon number $n$ sharply increases by several orders of magnitude, and at $N>N_c$ the photon number asymptotically approaches its limit value $(\Omega/\gamma)^2$ defined by Eq. (\ref{n2}). When  $N\gg N_c$  the number of pumping photons is less than the number of atoms and a further increase in $N$ does not affect the lasing process. This situation resembles a scenario described by Eberly \cite{Eberly} where subradiant spontaneous emission was considered in a system of $N$ atoms excited by a single photon. We call this regime, in which the number of photon does not depend on $N$ subradiant. It is interesting to follow the reverse transition from subradiant to superradiant regime by varying the pumping Rabi frequency $\Omega$ at fixed number of atoms $N$, as depicted in Fig. \ref{n(Om)}, where the number of photons is plotted as a function of $\Omega$. There exists some critical value of the pumping Rabi frequency $\Omega_c\approx\sqrt{\gamma g^2 N/3 k}$ which separates the two regimes of lasing. If we associate $\Omega$ with a number $n_P$ of pumping photons and compare it with the number of atoms $N$, then the ratio between the number of pumping photons and $N$ will define in which regime the laser works. Increasing $\Omega$ results in gradual increase of the number of photons, as long as $\Omega < \Omega_c$. When the pumping Rabi frequency reaches its critical value, i.e. at $\Omega = \Omega_c$, the number of photons drops several orders of magnitude and then slowly decreases with $\Omega$. To get more physical insight to this behavior, we show in Fig.\ref{Rho(Om)} the dependence of atomic populations on $\Omega$. In subradiant regime $\Omega < \Omega_c$ number of pumping photons $n_P$ is less than $N$, and the larger is $n_P$, the more atoms are involved into the lasing process giving rise to growing laser intensity. In this regime about 90\% of the atoms are in the two excited states. When $n_P$ becomes larger than $N$, strong pumping field results in sinchronization of fast Rabi oscillations on $\left|a\right\rangle\rightarrow\left|b\right\rangle$ and $\left|b\right\rangle\rightarrow\left|c\right\rangle$ transitions, the three atomic states are equally populated and the most energy is in atoms.

Figure \ref{PhaseDiag-Super} shows a ``phase diagram'' of the system where the solid (red online) line separates two different areas of parameters corresponding to superradiant (above the line) and subradiant (below the line) regimes.

Analysis of time behavior and the stability of the steady state solutions of Eqs. (\ref{rcc})-(\ref{b}) shows that at $N<N_c$ the superradiant solution is stable, while the subradiant one is unstable. At $N>N_c$ the situation is opposite: the subradiant solution becomes stable and the superradiant one becomes unstable. Thus the critical point $N=N_c$ is the instability point, which gives another indication to the presence of phase transition at this point. Figures \ref{Stab-Super} and \ref{Stab-Sub} illustrate different stability properties of the super and subradiant regimes. Phase portrait of the system at $N<N_c$ presented in Fig. \ref{Stab-Super} shows that in superradiant regime there is a single stationary point/attractor which is a stable node. In this regime the system gradually approaches its steady state that is stable.

Essentially different behavior is seen in Fig. \ref{Stab-Sub} showing phase portrait in the subradiant regime. The node at $\phi =\pi/2$ has lost its stability and transformed into a saddle point that repels slowly approaching phase trajectories, which then quickly reach the stable steady state defined by Eqs. (\ref{n2}) and $\Psi \approx 2\varphi+ArcTan[3\gamma/8\Omega]$. Stability analysis of the subradiant regime shows that first the field reaches the long-lived metastable state, and then, at some critical time, grows abruptly to stable steady state. In theory of phase transitions \cite{Strogatz2001} such a behavior is known as a phenomenon of critical slowing down.  \\ 

\emph{In summary}, a three-level ladder scheme with equally separated levels can generate a second harmonics of the pumping laser in either superradiant or subradiant manner, depending on the ratio between the number of atoms and the intensity of the pumping laser. Transition from one regime to another occurs abruptly, in phase-transition-like manner. In subradiant regime the steady state is also established in phase-transition-like manner, via critical slowing down. The laser field phase is locked to the phase of the driving laser. Details of the time evolution of this system will be published elsewhere.


\end{document}